\documentclass[aps,prb,twocolumn,showpacs]{revtex4}
\usepackage{graphicx}

\begin{document}

\title{Superconductivity at 22.3 K in SrFe$_{2-x}$Ir$_x$As$_2$}

\author{Fei Han, Xiyu Zhu, Ying Jia, Lei Fang, Peng Cheng, Huiqian Luo, Bing Shen and Hai-Hu Wen}\email{hhwen@aphy.iphy.ac.cn }

\affiliation{National Laboratory for Superconductivity, Institute of
Physics and Beijing National Laboratory for Condensed Matter
Physics, Chinese Academy of Sciences, P. O. Box 603, Beijing 100190,
China}

\begin{abstract}
By substituting the Fe with the 5d-transition metal Ir in
SrFe$_2$As$_2$, we have successfully synthesized the superconductor
SrFe$_{2-x}$Ir$_x$As$_2$ with T$_c$ = 22.3 K at x = 0.5. X-ray
diffraction indicates that the material has formed the
ThCr$_2$Si$_2$-type structure with a space group I4/mmm. The
temperature dependence of resistivity and dc magnetization both
reveal sharp superconducting transitions. An estimate on the
diamagnetization signal indicates a high Meissner shielding volume.
Interestingly, the normal state resistivity exhibits a roughly
linear behavior starting just above T$_c$ all the way up to 300 K.
The superconducting transitions at different magnetic fields were
also measured yielding a slope of -dH$_{c2}$/dT = 3.8 T/K near
T$_c$. Using the Werthamer-Helfand-Hohenberg (WHH) formula
$H_{c2}=-0.69(dH_{c2}/dT)|_{T_c}T_c$, the upper critical field at
zero K is found to be about 58 T. Counting the possible number of
electrons doped into the system in SrFe$_{2-x}$Ir$_x$As$_2$, we
argue that the superconductivity in the Ir-doped system cannot be
reconciled with the Co-doped case based on a simple rigid band
model, which should add more ingredients to understanding the
underlying physics of the iron pnictide superconductors.
\end{abstract} 
\maketitle

Since the discovery of superconductivity in LaFeAsO$_{1-x}$F$_x$
 \cite{Hosono}, the FeAs-based superconductivity has stimulated great
 interests in the fields of condensed matter physics and material
 science. In the ZrCuSiAs structure, the $T_c$ has been quickly promoted to 55-56
K\cite{Ren,cp} in fluorine doped or oxygen deficient oxy-pnictides
REFeAsO (RE = rare earth elements) and rare earth elements doped
fluoride-arsenide AeFeAsF (Ae = Ca,Sr) compounds\cite{xyzhu,cp2}.
Later on, in the system of (Ba,Sr)$_{1-x}$K$_x$Fe$_2$As$_2$ with the
ThCr$_2$Si$_2$-type structure (denoted as FeAs-122), the maximum
T$_c$ at about 38 K was discovered\cite{BaKparent,Rotter,CWCh}. This
FeAs-122 phase provides us a great opportunity to investigate the
intrinsic physical properties since large scale crystals can be
grown.\cite{Canfield} Furthermore, it was found that a substitution
of Fe ions with Co can also induce superconductivity with maximum
T$_c$ of about 24 K\cite{Sefat,XuZA}. Meanwhile, Ni substitution at
Fe site in BaFe$_2$As$_2$ has also been carried out with a T$_c$ of
20.5 K\cite{BaNiFeAs}. Very recently, superconductivity in Ru
substituted BaFe$_{2-x}$Ru$_{x}$As$_{2}$ was
found\cite{BaFe2-xRuxAs2}. This indicates that, the
superconductivity can be induced by substituting the Fe with not
only the 3d-transition metals, such as Co and Ni, but also the 4d
metal, like Ru. In this Communication, we report the successful
fabrication of the new superconductor SrFe$_{2-x}$Ir$_{x}$As$_{2}$
with a T$_c$ of about 22 K by replacing the Fe with the
5d-transition metal Ir. X-ray diffraction pattern (XRD),
resistivity, DC magnetic susceptibility and upper critical field
have been determined on this Ir-doped superconductor. Our discovery
here will add extra ingredients in understanding the underlying
physics in the iron pnictide superconductors.

The polycrystalline samples SrFe$_{2-x}$Ir$_x$As$_2$ were
synthesized by using a two-step solid state reaction
method\cite{xyzhu2}. Firstly, SrAs, FeAs and IrAs$_2$ powders were
obtained by the chemical reaction method with Sr pieces, Fe powders
(purity 99.99\%), Ir powders (purity 99.99\%) and As grains. Then
they were mixed together in the formula
SrFe$_{2-x}$Ir$_{x}$As$_{2}$, ground and pressed into a pellet
shape. All the weighing, mixing and pressing procedures were
performed in a glove box with a protective argon atmosphere (both
H$_2$O and O$_2$ are limited below 0.1 ppm). The pellet was sealed
in a silica tube with 0.2 bar of Ar gas and followed by heat
treatment at 900 $^o$C for 50 hours. Then it was cooled down slowly
to room temperature. A second sintering by repeating above process
normally can improve the purity of the sample.

The x-ray diffraction measurement was performed at room temperature
using an MXP18A-HF-type diffractometer with Cu-K$_{\alpha}$
radiation from 10$^\circ$ to 80$^\circ$ with a step of 0.01$^\circ$.
The analysis of x-ray powder diffraction data was done by using the
software of Powder-X,\cite{DongC} and the lattice constants were
derived (see below). The DC magnetization measurement was carried
out on a superconducting quantum interference device (SQUID)
magnetometer (Quantum Design MPMS-7T). In order to remove the effect
given by the possible residual magnetic field of the magnet, before
the measurement the system was degaussed. In this case the residual
magnetic field is limited below 1 Oe. The resistivity measurements
were done in a physical property measurement system (Quantum Design,
PPMS-9T) with magnetic fields up to 9$\;$T. The six-lead method was
used in the measurement on the longitudinal and transverse
resistivity at the same time. In this paper we report only the
resistivity data and leave other transport properties to be
published elsewhere. The temperature stabilization during the
measurement was better than $0.1\%$ and the resolution of the
voltmeter was better than 10$\;$nV.

\begin{figure}
\includegraphics[width=8cm]{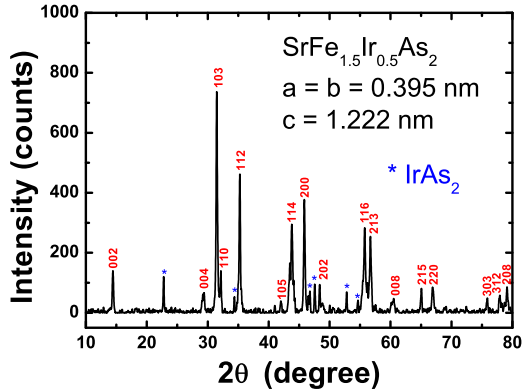}
\caption{(Color online) X-ray diffraction patterns for the sample
SrFe$_{1.5}$Ir$_{0.5}$As$_{2}$. Almost all main peaks can be indexed
by a tetragonal structure with $a = b = 3.95\mathrm{\AA}$ and $c =
12.22\mathrm{\AA}$. The asterisks mark the peaks arising from the
impurity phase (most probably the IrAs$_2$). } \label{fig1}
\end{figure}

In Figure.~\ref{fig1} we present the x-ray diffraction patterns of
SrFe$_{1.5}$Ir$_{0.5}$As$_{2}$. The main peaks can be indexed by a
tetragonal structure with $a=b=3.95\mathrm{\AA}$ and
$c=12.22\mathrm{\AA}$. In the parent phase
SrFe$_2$As$_2$,\cite{ParentSrFe2As2} the lattice constants $a$ (or
$b$) =3.92 $\AA$, $c$ = 12.36 $\AA$. Therefore by doping Ir into the
Fe site, the lattice constant $a$ shrinks a bit, while $c$ expands
slightly. This tendency is similar to the case of doping potassium
to the sites of Ba in Ba$_{1-x}$K$_x$Fe$_2$As$_2$, or substituting
the Fe with Ru in BaFe$_{2-x}$Ru$_{x}$As$_{2}$.\cite{BaFe2-xRuxAs2}.
There are still some small peaks which are coming from the secondary
phase, as marked by the asterisks. Further analysis indicates that
this tiny amount of impurity is most probably given by IrAs$_2$
since other peaks with asterisks can be indexed to the structural
data of IrAs$_2$. Concerning the very large Meissner shielding
volume as shown below, the XRD data here shows no doubt that the
superconductivity arises from the SrFe$_{1.5}$Ir$_{0.5}$As$_{2}$
phase. We should mention that the composition of Ir here gives only
the nominal value. A detailed analysis of the true composition on
the grains is under way. In addition, above the nominal doping level
of x=0.5, we found that the samples fabricated so far contained much
more impurity phases. A further refinement on the sample purity at a
high doping is strongly desired to illustrate the phase diagram of
the system.

\begin{figure}
\includegraphics[width=8cm]{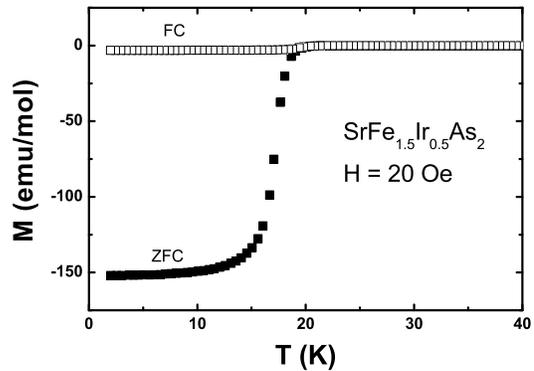}
\caption {(Color online) Temperature dependence of DC magnetization
for the sample SrFe$_{1.5}$Ir$_{0.5}$As$_{2}$. The measurement was
done under a magnetic field of 20 Oe in zero-field-cooled and
field-cooled modes. A strong Meissner shielding signal was observed
here.} \label{fig2}
\end{figure}

In Figure.~\ref{fig2} we present the temperature dependence of DC
magnetization for the sample SrFe$_{1.5}$Ir$_{0.5}$As$_{2}$. The
measurement was carried out under a magnetic field of 20 Oe in
zero-field-cooled and field-cooled processes. A clear diamagnetic
signal appears below 20.7$\;$K, which corresponds to the middle
transition temperature of the resistivity data. A very strong
Meissner shielding signal was observed in the low temperature
regime. Considering the shape of the sample (round disc with a
diameter of 5 mm and thickness of 1.3 mm), the Meissner shielding
volume is close to be full. We should mention that due to the
uncertainty in counting the demagnetization factor, it is difficult
to calculate the precise volume of the Meissner shielding. However,
the strong diamagnetization value is found to be among the strongest
magnetization (in emu/mol) found so far in the iron pnictide
superconductors, which signals a rather large volume of
superconductivity in the present sample.

\begin{figure}
\includegraphics[width=8cm]{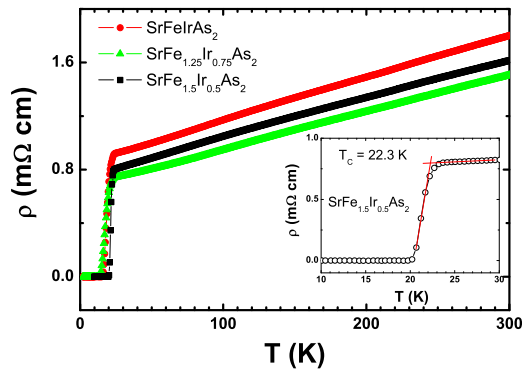}
\caption {(Color online) Temperature dependence of resistivity for
sample SrFe$_{2-x}$Ir$_{x}$As$_{2}$ (x = 0.5, 0.75, 1). The T$_c$
value here for the sample with x=0.5 was determined with a classical
method, i.e., the crossing point of the normal state background and
the extrapolation of the most steep transition part. A roughly
linear temperature dependence in the whole temperature region is
obvious.} \label{fig3}
\end{figure}

Figure.~\ref{fig3} shows the temperature dependence of resistivity
for samples SrFe$_{2-x}$Ir$_{x}$As$_{2}$  with x = 0.5, 0.75 and 1,
respectively. It was shown that the parent phase exhibits a sharp
drop of resistivity (resistivity anomaly as called so far) at about
205 K.\cite{Krellner} However in our superconducting samples, this
anomaly disappeared completely. The sample with nominal composition
x=0.5 shows a superconducting transition at about 22.3 K which is
determined by a classical method, i.e., the crossing point of the
normal state background and the extrapolation of the transition part
with the most steep slope. The transition width determined here with
the criterion of 10-90 $\%$ $\rho_n$ is about 1.8 K with $\rho_n$
the normal state resistivity (as marked by the dashed line Fig.3).
With higher doping (x=0.75 and 1.0) the transition temperature
declines slightly. As mentioned before, the samples with higher
doping levels (x=0.75 and 1.0) contain much more impurities,
therefore we are not sure whether this slight drop of
superconducting transition temperature is due to the chemical phase
separation or it is due to the systematic evolution of T$_c$ vs.
doping level. Interestingly, the normal state resistivity of the
superconducting sample x=0.5 shows a roughly linear behavior staring
just above T$_c$ all the way up to 300 K. It is certainly illusive
to know whether this linear behavior keeps going down to very low
temperatures when the superconductivity is suppressed by the strong
magnetic field. Further efforts are worthwhile to unravel whether
this linearity reflects an intrinsic feature of a novel electron
scattering. Since the sample with x=0.5 shows already a reliable
quality, we would believe that this linear temperature dependence of
resistivity is intrinsic and may posses itself of great importance.
More data are desired to clarify this interesting feature in the
normal state, together with the systematic evolution of the
superconducting properties by doping Ir.

\begin{figure}
\includegraphics[width=8cm]{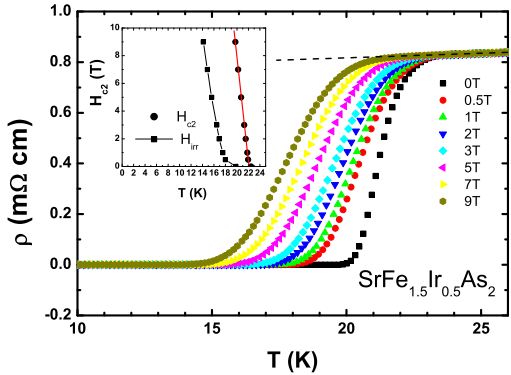}
\caption {(Color online) Temperature dependence of resistivity for
sample SrFe$_{1.5}$Ir$_{0.5}$As$_{2}$ at different magnetic fields.
The dashed line indicates the extrapolated resistivity in the normal
state. One can see that the superconductivity seems to be robust
against the magnetic field and shifts slowly to the lower
temperatures. The inset gives the upper critical field determined
using the criterion of 90\%$\rho_n$. A slope of -dH$_{c2}$/dT = 3.8
T/K near T$_c$ is found here. The irreversibility line H$_{irr}$
taking with the criterion of 0.1\% $\rho_n$ is also presented in the
inset.} \label{fig4}
\end{figure}

In Fig.4 we present the temperature dependence of resistivity under
different magnetic fields. Just as many other iron pnictide
superconductors, the superconductivity is very robust against the
magnetic field in the present sample. We used the criterion of
$90\%\rho_n$ to determine the upper critical field and show the data
in the inset of Fig.4. A slope of -dH$_{c2}$/dT = 3.8 T/K can be
obtained here. This is a rather large value which indicates a rather
high upper critical field in this system. In order to determine the
upper critical field in the low temperature region, we adopted the
Werthamer-Helfand-Hohenberg (WHH) formula\cite{WHH}
$H_{c2}=-0.69(dH_{c2}/dT)|_{T_c}T_c$. Taking $dH_{c2}/dT)|_{T_c}$= -
3.8 T/K and $T_c$ = 22.3 K, we have $H_{c2}(0)$ = 58 T. This
indicates that the present Ir-doped sample has also a very large
upper critical field, as in K-doped\cite{WangZSPRB} and Co-doped
samples.\cite{Jo} Very recently the high upper critical fields, as a
common feature in the iron pnictide superconductors, were
interpreted as due to the strong disorder effect.\cite{HighHc2}

The superconductivity induced by doping Co in
(Ba,Sr)(Fe$_{1-x}$Co$_x$)$_2$As$_2$ can be understood that electrons
are introduced into the system, which suppresses the
antiferromagnetic order.\cite{DaiPC} It was reported that the
maximum superconducting transition $T_c$ in the Co-doped system is
about 24 K which occurs at the doping level of x =
0.08.\cite{Canfield2,Fisher} Assuming the Co ions in this sample
have valence state of "3+", therefore the optimized
superconductivity takes place when 0.08/Fe electrons are doped into
the sample. When the doping level is about 0.16/Fe, it was found
that the superconductivity disappeared in the Co-doped
case.\cite{Canfield2,Fisher} In the present case with Ir-doping,
this scenario seems inapplicable. The usual valence state of Ir
element in an oxide is $4+$, the superconducting sample with x=0.5
in SrFe$_{2-x}$Ir$_x$As$_2$ corresponds to a doping level of
0.5-electron/Fe, this is already far beyond the value of 0.16/Fe in
achieving superconductivity with Co-doping. Therefore the
superconductivity in the Co-doped and Ir-doped cases cannot be
reconciled by a consideration based on the rigid band model. A band
structure calculation to such a high doping level in the Ir-doped
case is strongly recommended. In addition, the outer shell of
Ir$^{4+}$ ion has 5 electrons left, which is an odd number and is
different from the cases of Fe$^{2+}$ (6 electrons), Ni$^{2+}$ (8
electrons) and Co$^{3+}$ (6 electrons). Therefore it is a poorly
footed argument that the superconductivity occurs only in the system
with even electrons in the outer shell of the ions. Furthermore the
electrons of the 5d transition metals have higher itinerancy than
the 3d ones. In a naive picture, it would suggest that the higher
itinerancy of the doped electrons in the Ir-doped sample may lead to
the superconductivity. All these interesting hypothesis concerning
the superconductivity in our Ir-doped samples put more ingredients
to understanding the underlying physics of the pnictide
superconductors.

In summary, superconductivity with T$_c$ = 22.3 K has been observed
in SrFe$_{1.5}$Ir$_{0.5}$As$_2$. The normal state resistivity
exhibits a roughly linear behavior starting just above T$_c$ all the
way up to 300 K. This may reflect a novel scattering mechanism in
the normal state.  The superconductivity is rather robust against
the magnetic field with a slope of -dH$_{c2}$/dT = 3.8 T/K near
T$_c$. Using the Werthamer-Helfand-Hohenberg (WHH) formula
$H_{c2}=-0.69(dH_{c2}/dT)|_{T_c}T_c$, we got the upper critical
field at zero K of about 58 T. Based on the estimate on the high
electron density (0.5-electron/Fe) in the Ir-doped superconducting
sample, we argue that the superconductivity induced by doping Ir and
Co cannot be reconciled with the simple rigid band model.

This work is supported by the Natural Science Foundation of China,
the Ministry of Science and Technology of China (973 project:
2006CB601000, 2006CB921802), the Knowledge Innovation Project of
Chinese Academy of Sciences (ITSNEM).

\end{document}